\documentstyle[12pt]{article}
\newcommand{\doublespace}{\renewcommand{\baselinestretch}{1.75}
   \Large\normalsize}

\renewcommand{\ref}[1]{\raisebox{.6ex}{[#1]}}

\newcommand{\be}{\begin{equation}}
\newcommand{\ee}{\end{equation}}
\newcommand{\r} {{\bf r}}
\newcommand{\k} {{\bf k}}

\begin{document}

\doublespace

\title{ Double-Tip STM for Surface Analysis }

\author{ Q. Niu, M. C. Chang, and C. K. Shih  \\
University of Texas, Austin, TX  78712\\  }

\maketitle

\begin{abstract}
We explore the possibility of using a double-tip STM to probe the single
electron Green function of a
sample surface, and describe a few important applications:  (1) Probing
constant energy surfaces in $\k$-space by ballistic transport; (2)
Measuring scattering phase shifts of defects; (3) Observing the transition from
ballistic
to diffusive transport to localization; and (4) Measuring inelastic mean free
paths.


\vspace{5. cm}
To appear in Phys.~Rev.~B (Brief report)
\end{abstract}

\newpage

The single-tip scanning tunneling microscopy (STM) has become a major tool for
surface analysis \cite{STM}.  However, its use has been limited to probing
static
properties of electronic systems  such as the local density of states on or
near
sample surfaces.  In addition, it lacks the $\k$ resolution to enable one to
determine the
energy dispersion in band structures.  Transport properties are also out of the
reach of the single-tip STM, except for the BEEM configuration used to probe
ballistic
transport across a metal film \cite{BEEM}.  In this letter, we explore the
possibility of realizing a
double-tip STM, and describe a few important applications.

In a typical double-tip experiment, electrons are emitted from one tip,
and propagate through the sample, some of which are picked up by the other tip.
Naturally, the
propagator or the Green function of an electron in the sample is involved.
Since
all single particle properties of the system can be derived from the Green
function,  one expects
that a lot more information about the sample surface and nearby region can be
learned from an experiment
using a pair of tips than from using a single tip.  Some important applications
are:  (1)
deducing useful information about the band structure of surface states,  (2)
measuring scattering
phase shifts of surface defects, (3) Observing transition from ballistic to
diffusive transport to
localization, and (4) Measuring inelastic mean free paths.

Shown in Fig. 1 is a schematic experimental set up.  The sample is assumed to
be large
enough to have a well-defined chemical potential $\mu_0$.  Voltages $V_1$ and
$V_2$ are
applied to the tips relative to the sample, and electric currents $I_1$ and
$I_2$ from
the tips to the sample are measured.  Like the ballistic energy electron
microscope (BEEM),
this is a three terminal set up \cite{BEEM}.  Unlike the BEEM, here the
tip-sample
separations, the tip biases (V1 and V2), and the tip locations ($\r_1$ and
$\r_2$) are
controlled independently.   Direct junction conductances at $\r_1$ and $\r_2$
are defined as
$\sigma_i = {\partial I_i\over\partial V_i}$, and are given to leading order in
tunneling
rate as \cite{junction}
\be \sigma_i={2\pi e^2\over \hbar}\Gamma_i \rho(\r_i,\mu_i)\ee
where $\rho(\r_i,\mu_i)$ is the local density of states of the sample at the
chemical
potential of tip-$i$, and $\Gamma_i$ describes tip-sample couplings as well as
the density of
states of the tips (which are routinely measured in single-tip STM
experiments).  Up to
order $\Gamma_1\Gamma_2$, there are also processes of coherent tunneling of an
electron from one tip to the other through the sample, which can be measured
through
the transconductance defined by $\sigma_{21} \equiv {\partial I_2 \over
\partial V_1}$.
The transition rates can easily be accounted for by the Fermi golden rule using
second order
transition matrix elements \cite{Feynman}, yielding \be
\sigma_{21}=\Gamma_1\Gamma_2 {2\pi e^2\over \hbar}
|G(\r_1,\r_2;\epsilon=\mu_1)|^2,
\ee
where $G(\r_1,\r_2;\epsilon)$ is the retarded Green function of the
sample, for non-interacting electrons at zero temperature.

In the presence of electron-electron and/or electron-phonon interactions in the
sample, the Fermi golden rule gives the same result if $\mu_1=\mu_2$ and if the
sample
is nonsuperconducting.  For $\mu_1\ne\mu_2$, four-point Green functions of
the sample are involved to account for the inelastic scattering of the
tunneling electrons \cite {Chang}. For a superconducting sample, Andreev
processes can also
contribute to the transconductance within the same order \cite{Flatte}.
With the direct conductance and
transconductance measured at $\mu_1=\mu_2=\mu$, one can thus obtain information
about the retarded Green
function as shown below \be
|G(\r_1,\r_2;\epsilon=\mu)|^2={2\pi e^2\over \hbar}{\sigma_{21}\over
\sigma_1\sigma_2}\rho(\r_1,\mu)\rho(\r_2,\mu).
\ee
Since constant-current-STM-images trace constant contours of local density of
states,
the last two factors
can be treated as a normalization constant.

We next identify experimental parameters from
which such a measurement can be realized.  As a second order process, it is
clear that $\sigma_{21}$
is a weak signal to detect.  Taking BEEM as a reference, for a total tunneling
current of 1-10 nA,
the detecting limit is about 0.1-1 pA for ballistically transported electrons,
corresponding to a factor
of $10^{-3} - 10^{-4}$.  However, it is possible to utilize a frequency lock-in
amplifier to boost
this limit to $10^{-4} - 10^{-6}$.  For the sake of argument, we set a
conservative number of
$10^{-4}$ as a practical detection limit.  Considering a symmetric set-up such
that $\sigma_1 = \sigma_2
= \sigma$, from equation 3 one immediately finds that $\sigma_{21} /\sigma$ is
on the order of
${\hbar\over 2\pi e^2} \sigma$,  with additional factors determined by the
ratio between the Green
function and local density of states.  For a case of ballistic transport
through surface states (see
below),  $|G|^2/\rho^2$ is approximately $2\pi\over kr$, where $k$ is the wave
number at energy $\mu$,
and $r$ is the distance between the tips.  The closest tip-to-tip distance is
determined by the radius
of curvature and the aspect ratio of the tips.
Recent advances of
tip-fabrication techniques can reproducibly make high aspect ratio tips with a
radius of curvature on the
order of 50 - 100 \AA $\ $\cite{curvature}. It is thus conceivable to consider
operating a double-tip STM
in the range of  300 - 1000 \AA $\ $tip-to-tip distance.   This gives
$|G|^2/\rho^2$ on the order of one
percent.   Since $h/e^2 = 25$ k$\Omega$, one can immediately identify the
practical operation range of
tunnel junction resistance to be on the order of 1-10 M$\Omega$.  For the case
of an anisotropic Fermi
surface (as discussed below), $|G|^2/\rho^2$ is
on the order of unity in the same range of consideration
and the STM junction resistance can be as large as 1 G$\Omega$.
The most advantageous case would be a one dimensional
system, such as the fullerene nanotube, where the transconductance
does not depend on distance, and is of order unity. In the
following, we describe some
of the most important applications of the double-tip STM.

{\it -Ballistic Transport and Surface State Band Structure.   }
Consider a situation where surface states play a dominant role in electron
transport.
It is well known that the Green function for 2D free electrons is given by
\be -{im\over 2\hbar^2} H^{(1)}_0(kr)\approx -{im\over 2\hbar^2}
\sqrt{2\over  \pi k r } e^{i(kr-{\pi\over 4})}
\ee
where $k$ is the wave number, $r=|\r_2-\r_1|$ is the tip-tip distance, and
$H^{(1)}_0$
is a Hankel function of the first kind.  The right hand side is valid in the
asymptotic
region $kr>>1$, which is also the regime of our interest.  Therefore, the
transconductance decreases inversely with $r$ and is isotropic.  On an actual
crystalline
surface, the states are Bloch waves $e^{i\k \r} u_{nk}(\r)$ with band energies
$\epsilon_{n}(k)$.  It can be shown that for large $kr$,
the Green function can be approximated as \cite{Chang}
\be
G(\r_1,\r_2;\epsilon) \approx -iu_{n\k_c}(\r_1) u_{n\k_c}^*(\r_2)
\left(2\pi {\partial \epsilon_{n}\over
\partial k_\parallel} {\partial^2 \epsilon_{n}\over \partial
k_\perp^2} r \right)^{-1/2} e^{i[\k_c(\r_2-\r_1)-{\pi\over 4}]} \label{eq:
Green}
\ee
where $k_\parallel$ and $k_\perp$ are the components of $\k$ in the directions
parallel and perpendicular to $\r_2-\r_1$, respectively.
The partial derivatives are evaluated at the point $\k_c$ on the energy surface
where the normal vector ${\partial \epsilon_{n}\over
\partial \k}$ (group velocity) points in the direction of $\r_2-\r_1$.
In the following analysis we assume for simplicity that there is only a single
or single
dominant $\k_c$.\cite {kc's}

Like the free electron case, the transconductance has an overall inverse $r$
dependence.
However, two interesting new features appear for a crystal: (1) The
transconductance is
modulated by the magnitude squared of the Bloch functions.  (2) There is also
an overall
orientational dependence from the factors involving the partial derivatives of
the band
energy.  In Fig.2, the transconductance for a square lattice is
plotted at an energy close to a nested energy surface, showing an inverse $r$
dependence and
a pronounced anisotropy.  Bloch function modulation has
been removed by averaging over a unit cell.  Since the critical point $\k_c$
runs
through the energy surface as one changes the orientation of $\r_2-\r_1$, it is
possible to
reconstruct the constant energy contour of the surface band structure for the
filled and empty states
using the mapped-out $|G|^2$ .  The oscillatory modulation of the
transconductance should
also tell us the shape of the Bloch waves for each $\k_c$.  The energy
resolution of the tunneling
measurement is practically limited only by $k_BT$.  Sub-meV resolution is
routinely obtained \cite
{Hess}.  In comparison, angle-resolved photoemission (ARPES) can map-out
k-dependent band
structure only for the filled states and its resolution is currently limited to
about 15 meV both by the
spectrometer and the photon source.  A fourth power dependence of the
signal-to-noise ratio on
the demanded resolution makes it very difficult for ARPES to achieve
a finer energy resolution.

{\it -Phase shifts from a surface defect.  }
One can move the tips near (but still far compared to the wave length) to a
surface defect and
observe how it scatters the electrons by an interference effect.
An electron may propagate freely from tip1 to tip2, or it may propagate to the
defect and be
scattered to tip2.  The superposition of these processes can give rise to a
modulation of
$|G|^2$, when tip2 is moved around relative to tip1 and
the defect.  To illustrate the point, consider the 2D free electron model
again.  If
the scattering is dominated by the s-wave channel, the interference pattern
will consist
of curves of constant path-length-difference $r_1+r_2-r$, where $r_i$ is the
distance from the defect to tip-$i$.   Clearly, these are hyperbolic curves,
with the
positions of tip1 and the defects being the two focal points.  Quantitatively,
we have
\cite{Chang}
 \be
|G|^2/|G_0|^2\approx 1+\sqrt{8r\over\pi k r_1 r_2} \sin \theta
\cos (k(r-r_1-r_2)-{\pi\over 4}-\theta)
\ee
where $\theta$ is the phase shift of the defect.  The phase shift can thus be
determined
by the strength and the position of the modulation.

In the presence of additional scattering
channels, the above formula is generalized as
\be
|G|^2/|G_0|^2\approx 1+\sqrt{8r\over\pi k
r_1 r_2}
\sum_{m=-\infty}^\infty \cos (m\alpha)\sin \theta_m  \cos
(k(r-r_1-r_2)-m\pi-{\pi\over
4}-\theta_m) \ee
where $\theta_m$ is the phase shift for the $m$th angular momentum.  Now, the
interference pattern also depends explicitly on the angle $\alpha$ between the
vectors from
the defect center to the tips.

 On an actual solid surface, where defects scatter Bloch waves,
the above results should remain valid as long as the energy
surface is fairly isotropic.  In the general situation, one has to use
the scattering matrix $T_{\k_{c1},\k_{c2}}$, where $\k_{c1}$ is a Bloch wave
vector corresponding to a group velocity in the direction from tip1 to the
defect, and $\k_{c2}$ corresponds to a group velocity in the direction
from the defect to tip2.  The transconductance should behave like
\be
 |G|^2=|G_0(\r_1,\r_2)+G_0(\r_1,0)T_{\k_{c1},\k_{c2}}G_0(0,\r_2)|^2
\ee
where $G_0$ is the Green function in the absence of scattering.  The asymptotic
form of
this result at large distances can be obtained by using (5).

A single-tip STM can only see defects having a substantial disturbance of the
local density
of states on a sample surface.  The double-tip STM should be able to detect the
`mines'
buried fairly deep under a surface though the `radar' of scattering
interference.
Take again the free electron model (now 3D) to illustrate the point.  In terms
of the
positions of tip2, the contours of constant path length difference are now the
intersections of the sample surface with hyperbolic surfaces whose focal points
are at
tip1 and the defect center.  The defect can then be located by the
characteristics of
the interference pattern, e.g. it must lie under the symmetry axis of the
pattern.

{\it -Transition from Ballistic to diffusive transport to localization.  }
In the absence of defects, the transconductance measures ballistic electron
transport
between the tips.  As seen above, for surface states, this is signified by an
inverse $r$ dependence,
modulated by  an angular dependence if the energy surface is not circular.
When a defect is present
in the neighborhood of the two tips, scattering interference will occur.   The
interference
pattern will become more and more complicated if more defects are included in
the way
between the tips.  When the tips are moved apart by a few  elastic mean free
paths,
diffusive electron transport should begin to take place.
It can be shown that
$|G(\r_1,\r_2;\epsilon)|^2$ measures the time integral of the probability for a
wave  packet
of energy $\epsilon$ to move from $\r_1$ to $\r_2$ in a given time $t$
\cite{Chakravarty}.
In the diffusive
regime this probability goes as
\be
(\pi D t)^{-1}\exp ({-{r^2\over Dt}})
\ee
where $D$ is the diffusion constant.  Here, we should emphasize that this
formula only
describes the overall trend; statistical fluctuations can still persist in the
diffusive
regime in the form of the so called universal conductance fluctuations
\cite{fluctuation}.
The average transconductance
in this regime should then behave as  \be (\pi D)^{-1} E_1(r^2/Dt_c)
\ee
where $E_1(X)=\int_X^\infty {dx\over x} e^{-x}$ is the incomplete $\Gamma$
function, and
$t_c$ is a cutoff time beyond which elastic diffusive behavior ceases to occur.
 In the
absence of inelastic scattering, this cutoff time is given by $l_c^2/D$,
where $l_c$ is the Anderson localization length \cite{fluctuation}.
It is interesting to see
that the transconductance for $r<<l_c$ behaves like $(\pi D)^{-1} 2\ln
(l_c/r)$.
This slow fall off with distance is in sharp contrast with the behavior in the
ballistic
regime.  Also, the angular dependence should go away for a surface with square
or
hexagonal crystalline symmetry, because, as a second rank tensor, the diffusion
coefficient
cannot distinguish such point group symmetries from full isotropy.  Finally,
when $r$
is beyond the localization length, the Green function and the transconductance
should
drop exponentially.

{\it -Inelastic mean free path. }
In the above discussions, we have ignored electron-electron and electron-phonon
interactions in order to simplify the presentation.  It is well known that the
single
particle Green function will acquire a self energy with an imaginary part as
well as a real
part due to such interactions \cite{Mahan}.  The imaginary part pushes the
poles of the Green
function off the real axis, yielding an exponential decay of the Green function
in distance
$r=|\r_2-\r_1|$.  Physically, an electron tunneled in from tip1 may lose energy
by exciting
electron-hole pairs or emitting phonons as it travels in the sample, and become
unable to
tunnel out to tip2.  The typical length scale over which such a process occurs
defines
the inelastic mean free path, and is given by the decay length of the Green
function.
Therefore, the inelastic mean free path and its energy dependence may be
measured by
observing how the transconductance decays with tip-tip distance, and how the
decay length
varies with $\mu-\mu_0$.  For $\mu-\mu_0$ above the Debye energy (tens of meV,
hot
electrons), phonon emission dominates the inelastic processes.  At lower energy
differences,
electron-hole pair excitations become dominant.  In a Fermi liquid without
disorder, the
mean free path goes as  $h E_F v_F/(\mu-\mu_0)^2$.  A different energy
dependence has been predicted for
non-Fermi liquid systems.  The ability to measure the energy-dependent mean
free path using the
double-tip method should have important impact on this issue

{\it -Summary and Discussion.  }
Because a double-tip STM can probe the all important single particle Green
function of a
sample, it has the potential of becoming an extremely useful new tool in
surface analysis.
We have identified key experimental parameters for such measurements.  We have
further described some basic
applications of a double-tip STM: (1) Probing $\k$ resolved band structure of
surface states, as
well as the shape of Bloch functions; (2) Measuring scattering phase shifts or
amplitude  of surface
defects; (3) Observing transition from ballistic to diffusive transport to
localization,
and (4) Measuring
inelastic mean free paths.

   Apart from these applications, one can consider applying a magnetic field to
observe the  cyclotronic motion of electrons in the semiclassical regime
\cite{cyclotron}, or
to possibly probe some properties of a quantum Hall system in a strong magnetic
field \cite{edge}.  One can also consider mapping out the quasi-particle band
structure of a superconductor through ballistic transport or gap structure
through
Andreev reflections as proposed in Ref.\cite{Flatte}.

\newpage
Figure Captions:

Fig.1 	Schematic diagram of the double-tip STM experimental set-up.
Tip 1 is biased at $V_1$ and tip 2 at $V_2$ relative to the sample.  $I_1$ and
$I_2$ are also measured
relative to the sample.  When $\mu_1>\mu_2$,  $I_2$ contains a transconductance
component due to the
co-tunneling process.

Fig.2  The distance and angular dependence of the transconductance for a
surface with square
symmetry.  $X$ and $Y$ are the coordinates of $\r_2-\r_1$
($100\sim 600$ lattice constants) on a sample surface.
The vertical axis
is the square of the Green function normalized by the local density of states,
which is
the same  as the transconductance normalized by the junction conductances up to
a constant factor.
The corresponding  energy surface (nearly nested) is shown on top of the
figure.


\begin{thebibliography}{99}
\bibitem{STM}  G. Binnig, H. Rohrer, Ch. Gerber, and E. Weibel, Phys. Rev.
Lett. 40, p178
(1982).
\bibitem{BEEM} W.J. Kaiser and L.D. Bell, Phys. Rev. Lett. 60, p1406
(1988); for a review, see L.D. Bell, W.J. Kaiser, M.H. Hecht, and L.C. Davis,
Methods of
Experimental Physics vol. 27, Scanning Tunneling Microscopy, edited by J.A.
Stroscio and W.J.
Kaiser, Academic Press 1993.
\bibitem{junction} J. Tersoff and D. R. Hamann, Phys. Rev. Lett. 50, p1998
(1983); Phys.
Rev. B 31, p805 (1985).
\bibitem{Feynman} R. P. Feynman and A. R.
Hibbs, Quantum mechanics and path integrals (McGraw-Hill, New York, 1965) Ch.
6.
\bibitem{Chang} M. C. Chang, Q. Niu and C. K. Shih, unpublished.
\bibitem{Flatte} J. M. Byers and M. E. Flatte,
to be published in Phys. Rev. Lett. Their result (Eq.(9) in preprint)
reduces to our Eq.(2) in the absence of pairing of electrons, but
their derivation was based on a Hamiltonian which is quadratic
in electron creation and destruction operators.
\bibitem{curvature} A.D. Kent, T.M. Shaw, S. von Moln\'ar, and D.D. Awschalom,
Science, 262,
1249, (1993).
\bibitem{kc's}
When there is more than one
critical point on the energy surface, the contributions from all the critical
points
must be included as a sum in the evaluation of $G$, and interference effect may
arise
if different terms are about the same in magnitude.
\bibitem{Hess} H.F. Hess, R.B. Robinson, and J.V. Waszczak, Phys. Rev. Lett.
64, 2711, (1990).
\bibitem{Chakravarty} S. Chakravarty and A. Schmid, Phys. Rep. 140, p193
(1986).
\bibitem{fluctuation} P. A. Lee and T. V. Ramakrishnan, Rev. Mod. Phys.
					57, 287 (1985).
\bibitem{Mahan} G. D. Mahan, Many-Particle Physics, 2nd ed. (Plenum, New York,
1990).
\bibitem{cyclotron}  For recent reviews, see A. M. Duif, A. G. M. Jansen, and
P. Wyder,
	J.Phys. Condens. Matter 1, 3157 (1989); and C. W. J. Beenakker, H. van Houten,
and B. J.
 van Wees, Europhys. Lett. 7, 359 (1988).
\bibitem{edge}  J. Kinaret, Y. Meir, N. Wingreen, P. A. Lee, and X. G. Wen,
Phys. Rev. B 46, 4681
(1992).


\end{thebibliography}
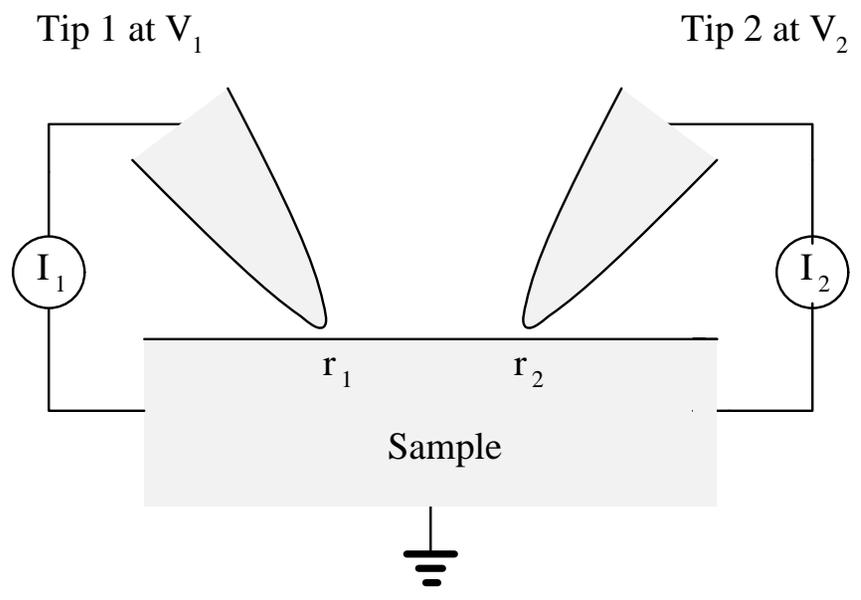
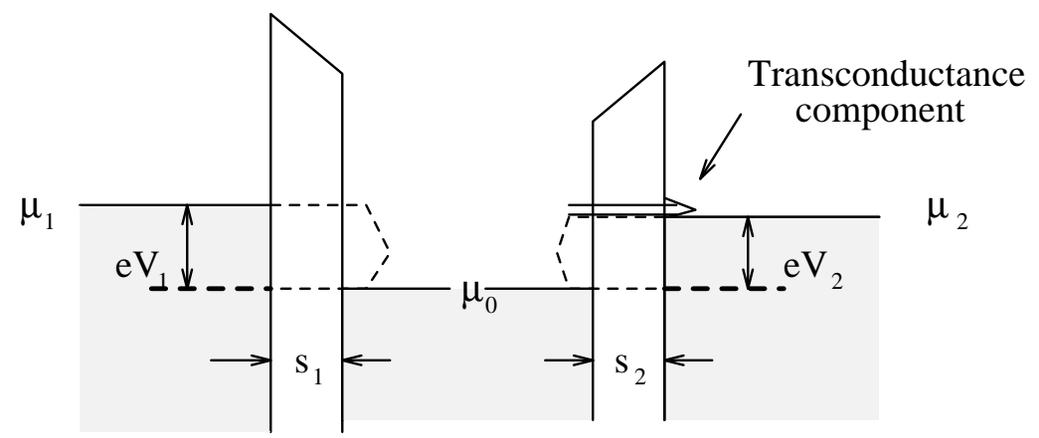
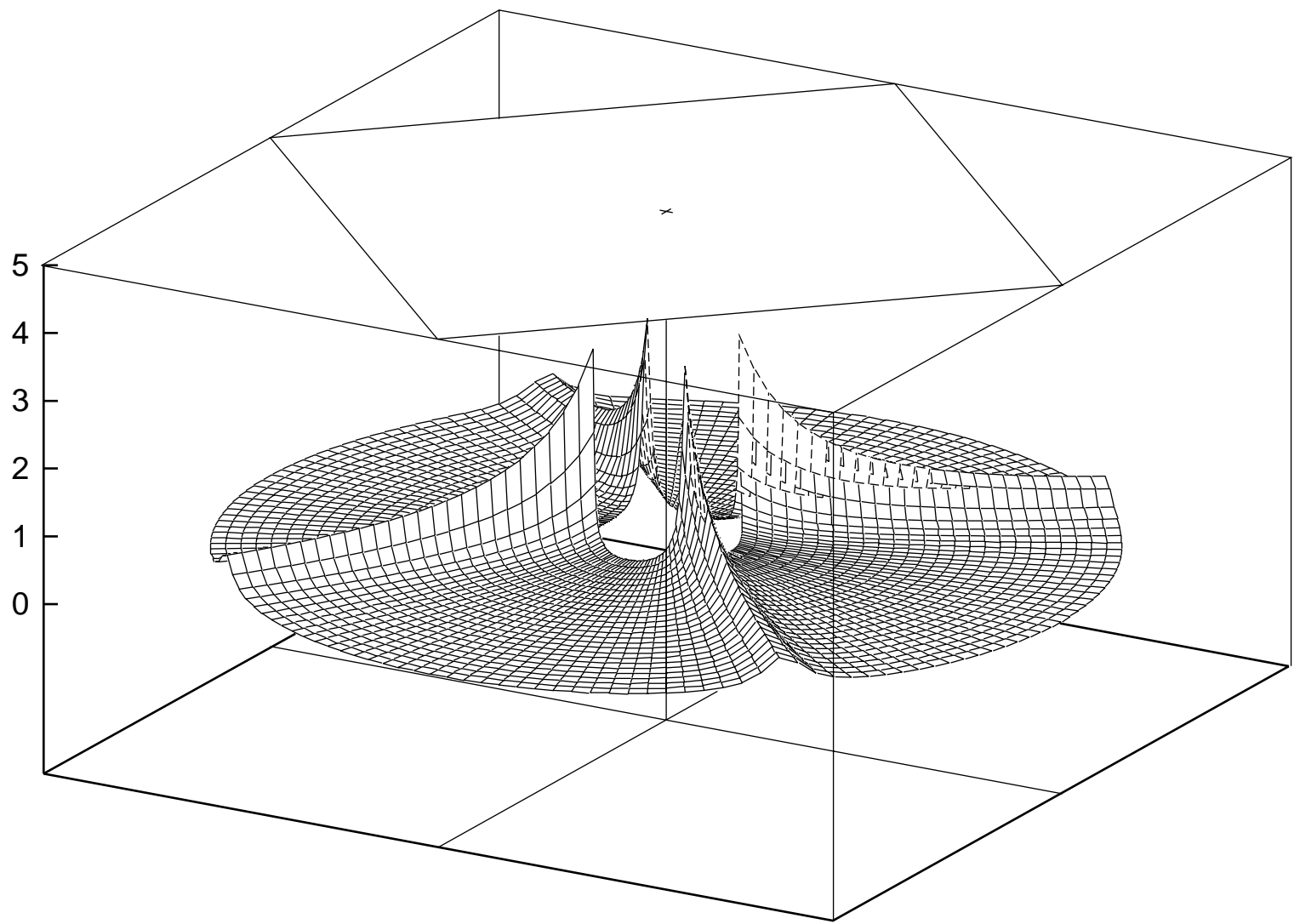
\end{document}